\documentclass[10pt,twoside,onecolumn]{IEEEtran}

\usepackage {rotating,cite,calc,color,eepic,marvosym,pxfonts,txfonts}
\usepackage {graphicx}
\usepackage{ifsym}
\usepackage{epstopdf}
\usepackage{algorithm}
\usepackage{algorithmicx}
\usepackage{algpseudocode}
\usepackage[font = footnotesize]{caption}
\usepackage{authblk}
\usepackage{amssymb,epsfig,tabularx,delarray,enumerate,graphicx,array,cite,xcolor,subcaption}

\title{Wavelength Assignment in Hybrid Quantum-Classical Networks}

\author[1,2,*]{Sima Bahrani}
\author[1]{Mohsen Razavi}
\author[2]{Jawad A. Salehi}
\affil[1]{School of Electronic and Electrical Engineering, University of Leeds, Leeds, LS2 9JT, UK}
\affil[2]{Electrical Engineering Department, Sharif University of Technology, Tehran, Iran}
\affil[*]{si.bahrani@gmail.com}
\begin{document}
	
	\maketitle

\begin{abstract}
Optimal wavelength assignment in dense-wavelength-division-multiplexing (DWDM) systems that integrate both quantum and classical channels is studied. In such systems, weak quantum key distribution (QKD) signals travel alongside intense classical signals on the same fiber, where the former can be masked by the background noise induced by the latter. Here, we investigate how optimal wavelength assignment can mitigate this problem. We consider different DWDM structures and various sources of crosstalk and propose several near-optimal wavelength assignment methods that maximize the total secret key rate of the QKD channels. Our numerical results show that the optimum wavelength assignment pattern is commonly consisted of several interspersed quantum and classical bands. Using our proposed techniques, the total secret key rate of quantum channels can substantially be improved, as compared to conventional assignment methods, in the noise dominated regimes. Alternatively, we can maximize the number of QKD users supported under certain key rate constraints.
\end{abstract}

\section{Introduction}
\label{Sec:Intro}
Quantum networks are no longer a physicist's fantasy but the emerging reality of today's complex communications world. With the first quantum satellite in orbit \cite{ChinaSatQKD,ChinaSatTelep}, and the launch of the 2000-km-long Beijing-to-Shanghai quantum key distribution (QKD) network \cite{China_BeiShang}, quantum technologies have reached a new milestone in supporting multiple users at long distances. This trend is boosted by various national and regional programmes in the UK, the European Union, the States, and far east Asia, that aim at bringing the technology to the doorsteps of the end users. While the emerging quantum communications technologies would offer future-proof security for our data exchange, they are not going to replace the vast investment in high data-rate communications. In fact, any commercially sensible solution for quantum networks would rely on its integration with classical infrastructure \cite{Telcordia_1550_1550, Telcordia_1550_1310,qi2010feasibility,Shields.PRX.coexist,patel2014quantum,CVQKD-DWDM}. In this paper, we address one of the problems that arise in such integrated networks. We look at a dense-wavelength-division-multiplexing (DWDM) setup in which multiple QKD channels are multiplexed with several data channels \cite{crosstalk2016,bahrani2016optimal}, and will investigate how wavelength assignment in such a setup can affect the amount of crosstalk, and consequently the performance of QKD channels.

From the first proposed QKD protocols \cite{BB_84} up until now, the field has seen considerable development and progress. Various QKD protocols and different techniques to address their practical issues have been proposed in the literature \cite{ekert1991quantum,Lo:Decoy:2005,ma2007quantum,lo2012measurement,panayi2014memory,piparo2015measurement}. For instance, one of the main advancements in the implementation of QKD is the decoy-state technique, which relaxes the need for an ideal single-photon source in QKD systems. On the other hand, there has been significant enhancement in reach and performance of point-to-point QKD links. The successful demonstration of measurement-device-independent QKD over $404~{\rm km}$ \cite{yin2016measurement} has proven the feasibility of running QKD over long optical fibers, although at a very low rate. To cover longer distances, at a high key-exchange rate, and to support multiple users, QKD must be implemented over large-scale networks.

The initial steps for the implementation of QKD in a network setting have been carried out successfully \cite{secoqc,Sasaki:TokyoQKD:2011, Wang:14, MDInetwork}. The European project SECOQC and the Tokyo QKD network each demonstrated a small quantum network with mesh topology. To extend these examples, the first generation of QKD networks are expected to rely on the key exchange in a trusted-note architecture. That is, in order to overcome distance limitations of QKD, a set of trusted nodes, at the core of the network, can be used to link the two end nodes. The most recent example of such networks is the developing link between Beijing and Shanghai with 32 trusted nodes along the way. 

Regardless of the topology, one major requirement in the widespread development of QKD networks is their integration with the existing fiber-optic classical networks. This is not limited to the current developing QKD networks, but also next generations of quantum networks should address this issue due to cost efficiency considerations. In such hybrid networks, weak quantum signals should travel alongside intense classical ones. The latter, in this scenario, would produce some background noise, e.g., Raman scattering and adjacent channel crosstalk, which will enter the quantum receivers. 

In order to reduce the background crosstalk in hybrid quantum-classical setups, several methods have been proposed. For instance, filtering methods in frequency and time domains have been used to suppress the crosstalk \cite{Shields.PRX.coexist,patel2014quantum}. Another useful approach is the control of launch power of data channels such that it satisfies the receiver sensitivity \cite{Shields.PRX.coexist,patel2014quantum}. Furthermore, it has been shown that orthogonal frequency division multiplexing (OFDM) can effectively reduce the crosstalk using an inherent optimal filtering \cite{crosstalk2016}.

In this paper, we propose optimal wavelength assignment as an additional method of crosstalk reduction in a DWDM link at the core of a hybrid network. The problem of optimal wavelength assignment in integrated quantum-classical DWDM systems with {\em one} quantum and several classical channels has been investigated for certain QKD systems \cite{CVQKD-DWDM}. However, the more general scenario, where multiple quantum and multiple classical signals are to be transmitted, has not been fully studied yet. Considering the shape of Raman spectrum, a conventional solution for this problem is the assignment of higher wavelengths to classical channels, and the lower wavelengths to quantum ones. Appropriate wavelength assignment under the constraint of having two separate quantum and classical bands has been investigated in an earlier work \cite{crosstalk2016}. However, it has been shown by the authors that optimal wavelength assignment does not necessarily follow this two-band form \cite{bahrani2016optimal}. Here, we investigate optimal wavelength assignment and propose several methods to approach it in different DWDM setups. We also show that the optimal wavelength assignment can improve the performance of QKD links effectively. 

In our optimization problem, we consider two particular scenarios. In the first scenario, our objective is to get the maximum aggregate key rate out of a fixed number of quantum channels in the presence of a number of classical channels. This scenario is relevant in the settings that all generated keys are to be consumed by our two end nodes. That is, the main point of multiplexing several QKD channels is to increase the total key rate. Another foreseeable scenario is that each quantum channel represents a different user. In such a case, we have to be able to guarantee a minimum key rate for all users. In such a setting, the optimized solution can maximize the number of users that can be supported. This may or may not coincide with the total maximum key rate as we show in this paper.


In the following, the hybrid quantum-classical DWDM system is described in Sec.~\ref{Sec:Sys}. In Sec.~\ref{Sec:Rate}, the key rate analysis is presented. The proposed wavelength assignment methods are described in Sec.~\ref{Sec:Opt}. We present our numerical results in Sec.~\ref{Sec:Num}, and conclude the paper in Sec.~\ref{Sec:Conc}.

\section{System Description}
\label{Sec:Sys}
Consider a DWDM link in the backbone of a quantum-classical network carrying several classical and quantum channels. We refer to the two end nodes of the link by Alice and Bob. We consider a general scenario where $M$ channels are assigned to the QKD usage, while $N$ forward classical channels (from Alice to Bob) and $N$ backward classical channels (from Bob to Alice) transmit classical data. As for the fiber link, we consider two cases of full-duplex over a single-mode fiber, and dual-fiber, as shown in Figs.~\ref{single_dual}(a) and (b), respectively. In the first case, we assume that each classical channel is equipped with optical circulators to enable the transmission of signals in both directions on the same wavelength. {This structure describes some of the existing, and probably future high-capacity, full-duplex DWDM systems with both classical and quantum links.} In the setup of Fig.~\ref{single_dual}(b), forward and backward data signals are transmitted via different fibers. {This structure is, for example, used in 100G systems. In this case, we assume that the classical and quantum signals are transmitted in the same direction on both fiber links. That is, in the forward link, QKD encoders are located on Alice's side, whereas for the backward link, they are on Bob's side. This assumption is based on the fact that the Raman noise generated in this case is smaller than the case of transmitting quantum and classical signals in the opposite directions \cite{Shields.PRX.coexist}, as we will explain later.

Let us introduce the notation we use for the employed wavelengths in our hybrid link. The set of available wavelengths in the system is denoted by $G=\{\lambda_{1},...,\lambda_{D}\}$. We denote the channel spacing of the DWDM system by $\Delta$. Furthermore, the set of wavelengths assigned to forward and backward classical channels are represented by $A=\{\lambda_{A_1},...,\lambda_{A_N}\}$ and $B=\{\lambda_{B_1},...,\lambda_{B_N}\}$, respectively. Note that sets $A$ and $B$, in both structures, could be overlapping. All classical signals are assumed to have equal launch power, denoted by $I$.  This launch power is assumed to be minimized, considering the receiver sensitivity, to meet a target bit error rate (BER). In the case of quantum channels, we introduce two wavelength sets: $U_{1}=\{\lambda_{q_1},...,\lambda_{q_k}\}$ represents the channels whose quantum signals travel from Alice to Bob, and $U_{2}=\{\lambda_{q_{k+1}},...,\lambda_{q_M}\}$ represents the channels whose quantum signals travel from Bob to Alice. In the dual-fiber structure, parameter $k$ will then represent the number of quantum channels on the forward link. Note that, $U_{1}$ and $U_{2}$ may also be overlapping. In the full-duplex system, $k=M$, and $U_2$ would be an empty set.

\begin{figure}[t]
\centering
\includegraphics[width=14 cm]{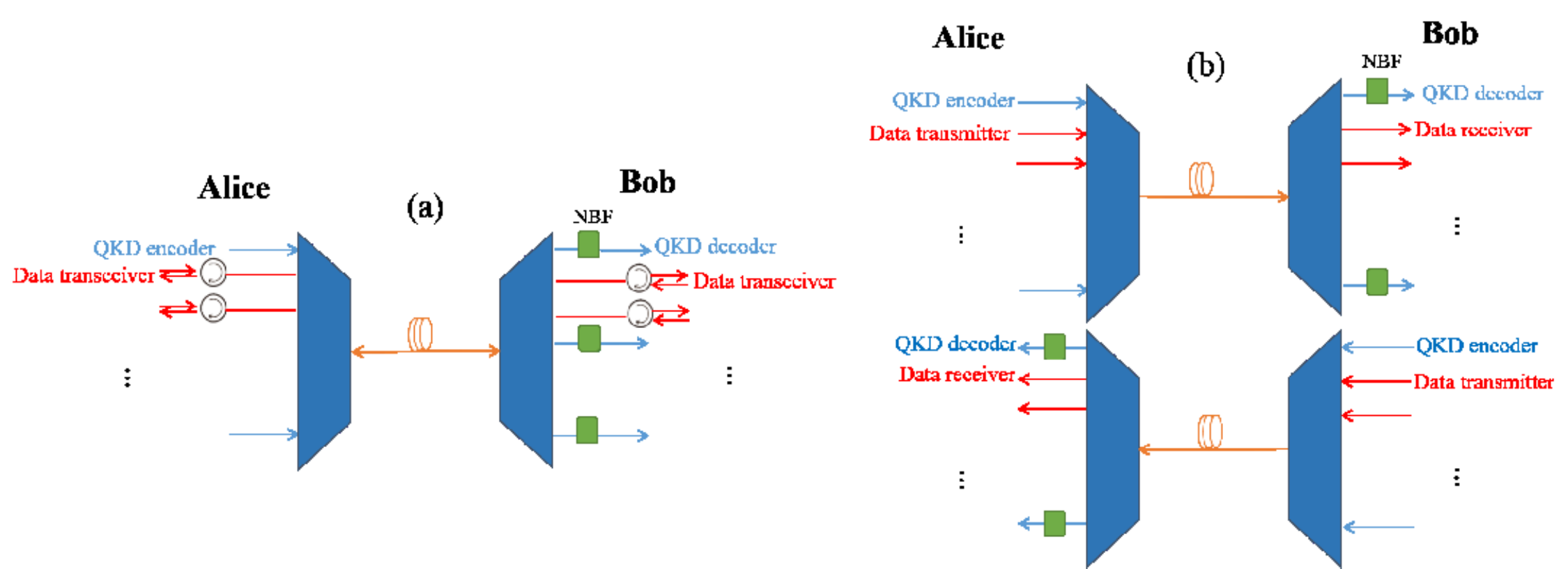}
\caption{Hybrid DWDM link with multiple quantum and classical channels: (a) A full-duplex DWDM system over a single-mode fiber. Each classical channel is equipped with optical circulators to enable the transmission of signals in both directions on the same wavelength. Quantum signals travel from Alice to Bob. (b) A dual-fiber DWDM system, where forward and backward data signals are transmitted via different fibers. Quantum signals travel along the same direction as the classical ones.} \label{single_dual}
\end{figure}  

In this paper, we use the BB84 protocol with time-bin encoding \cite{MXF:MIQKD:2012} for our QKD channels; see Fig. \ref{BB84}. {This method is mainly suitable for fiber channels.} We also use the decoy-state version of efficient BB84 \cite{Lo:EffBB84:2005} to allow for weak laser pulses to be used at QKD encoders. In the time-bin encoding, the qubits are encoded on the phase difference of two consecutive pulses, $r$ and $s$, generated by a Mach-Zehnder interferometer (MZI). The encoding phase, $\phi_{A}$, is chosen from one of the basis sets $\{0,\pi\}$, for ${X}$ basis, and $\{\pi/2,3\pi/2\}$, for ${Y}$ basis, randomly. At the QKD decoder, the decoding phase, $\phi_{B}$, of Bob's MZI is chosen randomly, from the set $\{0,\pi/2\}$. He then interferes the received $r$ and $s$ pulses by means of his MZI and infers the transmitted qubit by measuring the output pulses. 

\begin{figure}[t]
\centering
\includegraphics[width=3 in]{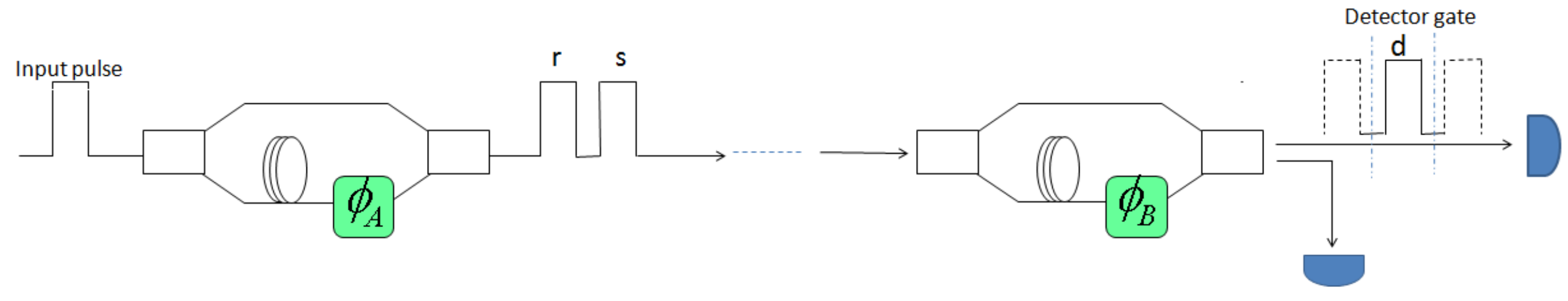}
\caption{Phase encoded (time-bin) QKD. Alice encodes her key bits by choosing a phase value, $\phi_{A}$, from one of the bases $\{0,\pi\}$ and $\{\pi/2,3\pi/2\}$. Each optical pulse passes through the MZI and produces two output pulses with the relative phase $\phi_{A}$. On the Bob side, another MZI is used to recombine $r$ and $s$ modes, followed by photodetection. Active phase and polarization maintenance is assumed to be in place.}\label{BB84}
\end{figure} 

The integration of classical and quantum signals on the same fiber results in certain problems that may affect the QKD operation. The main challenge is the background noise generated by the classical signals that reaches the quantum receivers. Two main sources of this crosstalk noise are the inelastic interactions in an optical fiber and nonideal operation of DWDM multiplexers and demultiplexers. In particular, the Raman scattering and the power leakage from adjacent channels have been shown to be the dominant sources of noise in such hybrid systems \cite{eraerds2010quantum}. In the following, the effect of these sources of background noise is described in more detail. We assume that our QKD decoders are equipped with narrow-band filters (NBF) to reduce such noises.


\subsection{Sources of crosstalk noise}

\subsubsection{Raman noise} 
Raman scattering occurs due to the inelastic photon-phonon interactions in an optical fiber. Because of its wide spectrum, Raman noise can easily leak into quantum channels. The Raman noise co-propagating with the data signal is referred to by forward scattering, whereas the Raman noise traveling in the opposite direction is known as backward scattering. Backward Raman scattering is known to be stronger than the forward one for typical fiber lengths \cite{Shields.PRX.coexist}. 
    
In the DWDM systems shown in Fig.~\ref{single_dual}, each classical signal induces a certain amount of Raman crosstalk noise at the receiver of each quantum link. Consider the quantum channel with wavelength $\lambda_{q_m}$ for $m=1,\ldots,M$, and let us calculate the amount of Raman noise induced by the $n$th, $n=1,\ldots,N$, data channel. The data channel can include signals traveling in the same direction as the quantum signals in channel $m$, or the opposite direction. Let us denote the wavelength of the former by $\lambda_{f_{n}}$, and the latter by $\lambda_{b_{n}}$. For instance, for the full-duplex structure, we have $\lambda_{f_{n}}=\lambda_{A_{n}}$ and $\lambda_{b_{n}}=\lambda_{B_{n}}$. In the dual-fiber case, if $m \leq k$, then $\lambda_{f_{n}}=\lambda_{A_{n}}$, otherwise $\lambda_{f_{n}}=\lambda_{B_{n}}$. There would be no backward classical channel in the dual-fiber case, hence $\lambda_{b_{n}}$ is not defined. With this notation, the forward Raman (FR) noise power corresponding to $\lambda_{f_{n}}$ and $\lambda_{b_{n}}$, respectively, for the $m^{th}$ quantum channel, is given by \cite{Shields.PRX.coexist,eraerds2010quantum}:
\begin{equation}
I^{\rm FR}_{nm}=I e^{-\alpha L}L \beta (\lambda_{f_{n}},\lambda_{q_{m}})\Delta \lambda
\label{Iforward}
\end{equation}
and, in the full-duplex case, the backward Raman (BR) noise is given by
\begin{equation}
I^{\rm BR}_{nm}=I \frac{(1-e^{-2\alpha L})}{2\alpha}\beta (\lambda_{b_{n}},\lambda_{q_{m}})\Delta \lambda,
\label{Ibackward}
\end{equation} 
where $\beta (\lambda_{d},\lambda_{q})$ is the Raman cross section (per fiber length and bandwidth) at wavelength $\lambda_q$ for a classical pump signal at wavelength $\lambda_d$. For our dual-fiber system, $I^{\rm BR}_{nm}=0$. In the above equations, $\alpha$, $L$, and $\Delta \lambda$ are, respectively, the fiber attenuation coefficient, the fiber length, and the bandwidth of the NBF in wavelength unit. {In this paper, for simplicity, we assume that $\alpha$ is constant across the employed wavelength grid. This is a good approximation for wavelengths within 1530~nm and 1565~nm in the C band, which is considered in our numerical results. It would be straightforward to use a wavelength dependent $\alpha$ if loss variations are substantial in the grid.} Figure \ref{raman_cross} shows measured Raman cross section, $\beta (1550 \rm{\ nm},\lambda)$, in a standard single mode fiber \cite{eraerds2010quantum}. As can be seen, the Raman cross section is slightly higher for wavelengths longer (Stokes regime) than 1550~nm than the ones below (anti-Stokes regime) it. That has resulted in a perception that perhaps the best way of allocating wavelengths to quantum and classical signals is to use the higher wavelengths for data channels and the lower ones for quantum. We refer to this solution as the conventional method, and will investigate how far or close it is to the optimum assignment we find in this paper.

\begin{figure}[t]
\centering
\includegraphics[width=3 in]{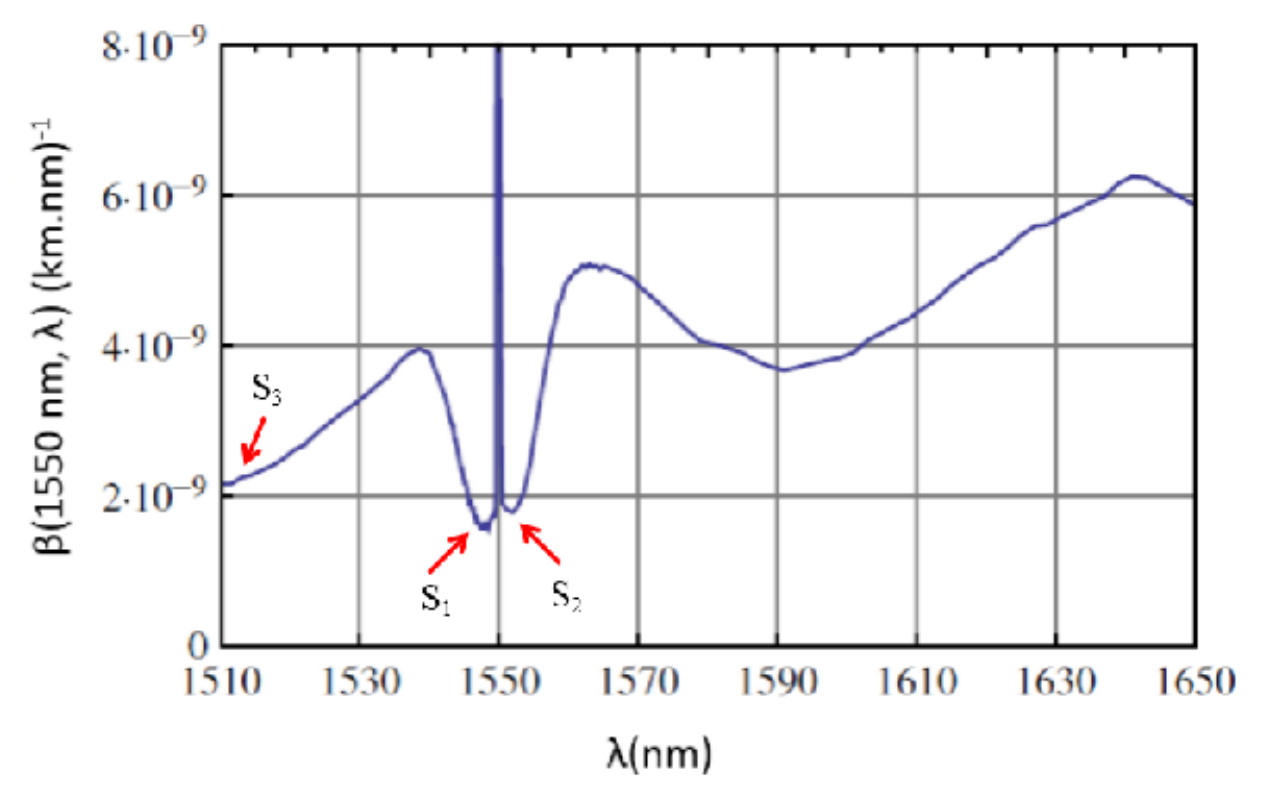}
\caption{Measured Raman cross section with respect to a central wavelength of 1550 nm in a standard single mode fiber \cite{eraerds2010quantum}.} \label{raman_cross} 
\end{figure}

The average Raman photon counts corresponding to $\lambda_{f_{n}}$ and $\lambda_{b_{n}}$, at the detectors of the $m^{th}$ quantum receiver is, respectively, given by
\begin{equation}
p^{\rm FR}_{nm}=I^{\rm FR}_{nm}\lambda_{q_m}T_{d}{\eta_d}/(2hc)=C^{f}\lambda_{q_m} \beta (\lambda_{f_{n}},\lambda_{q_{m}}),
\label{forward}
\end{equation}
and
\begin{equation}
p^{\rm BR}_{nm}=I^{\rm BR}_{nm}\lambda_{q_m}T_{d}{\eta_d}/(2hc)=C^{b}\lambda_{q_m} \beta (\lambda_{b_{n}},\lambda_{q_{m}}),
\label{backward}
\end{equation} 
where $C^{f}$ and $C^{b}$ are {\em wavelength-independent} parameters, given by
\begin{equation}
C^{f}= I e^{-\alpha L}L\Delta \lambda T_{d}{\eta_d}/(2hc),
\label{gamma_f}
\end{equation}
and
\begin{equation}
C^{b}= I \frac{(1-e^{-2\alpha L})}{2\alpha}\Delta \lambda T_{d}{\eta_d}/(2hc).
\label{gamma_b}
\end{equation}
In the above equations, $h$ and $c$ are, respectively, the Planck constant and the speed of light in the vacuum, $T_d$ is detectors' gate interval, and $\eta_d$ denotes their quantum efficiency. Note that, in the dual-fiber case, $p^{\rm BR}_{nm} = 0$. The (1/2) factor in above equations account for the loss in the passive decoder of Fig.~\ref{BB84}.


\subsubsection{Adjacent channel crosstalk}

The DWDM multiplexers and demultiplexers can also introduce some crosstalk noise because of their nonideal operation. Insufficient channel isolation can cause some power leakage from data channels to copropagating quantum channels. {Furthermore, there may be some back reflection from data signals transmitted in the opposite direction to quantum signals, into the quantum receivers.} In general, the power leakage from a classical channel into the two immediately adjacent channels is higher than that of the non-adjacent ones. Moreover, with the use of appropriate NBFs at the quantum receivers, the nonadjacent channel crosstalk can be suppressed effectively. In this paper, we only then consider the adjacent channel crosstalk. We denote the adjacent channel isolation of the DWDM module in dB by $\gamma_{a}$. Furthermore, the average value of the transfer function of the NBF at the passband of the adjacent channels is denoted by the coefficient {$g_{a}$}. Then, the power leakage corresponding to $\lambda_{f_{n}}$ and $m^{th}$ quantum channel can be expressed as
\begin{equation}
I^{\rm FC}_{nm}=\left\{
\begin{array}{cc}
g_{a}Ie^{-\alpha L}10^{(-\gamma_{a}/10)}& |\lambda_{f_n}-\lambda_{q_m}|=\Delta \\
0& |\lambda_{f_n}-\lambda_{q_m}|> \Delta 
\end{array}
.
\right.
\label{crossfor}
\end{equation}   
Similarly, in the full-duplex case, the power leakage corresponding to $\lambda_{b_{n}}$ at the $m^{th}$ quantum receiver is given by
\begin{equation}
I^{\rm BC}_{nm}=\left\{
\begin{array}{cc}
g_{a}I10^{(-\chi_{a}/10)}& |\lambda_{b_n}-\lambda_{q_m}|=\Delta \\
0& |\lambda_{b_n}-\lambda_{q_m}|> \Delta 
\end{array}
,
\right.
\label{crossback}
\end{equation} 
where $\chi_{a}$ represents the directivity of the DWDM multiplexer. The indices ``FC" and ``BC" represent ``Forward Crosstalk" and ``Backward Crosstalk", respectively. In the dual-fiber case, $I^{\rm BC}_{nm}=0$. The average photon counts corresponding to $I^{\rm FC}_{nm}$ and $I^{\rm BC}_{nm}$ are, respectively, obtained by  
\begin{equation}
p^{\rm FC}_{nm}=I^{\rm FC}_{nm}\lambda_{q_m}T_{d}{\eta_d}/(2hc),
\label{c_forward}
\end{equation}
and
\begin{equation}
p^{\rm BC}_{nm}=I^{\rm BC}_{nm}\lambda_{q_m}T_{d}{\eta_d}/(2hc).
\end{equation}

In our analysis, we neglect any crosstalk noise from quantum channels on each other. One possible source of such a noise can be the synchronization signals sent by the QKD systems. The rate at which such signals are sent is often very low and can be neglected. Alternatively, one/some of the classical channels can be used for time synchronization as well as other classical tasks that QKD systems require. Finally, we can also use time-multiplexing techniques to separate the quantum and synchronization signals on QKD channels. In the latter case, we assume that the time synchronization signals are transmitted simultaneously on all QKD channels to avoid any crosstalk noise on QKD signals.

\section{Key Rate Analysis}\label{Sec:Rate}
In this section, the secret key generation rate of the QKD links in the DWDM systems of Fig.~\ref{single_dual} is analyzed. We consider the $m^{th}$ QKD channel, as an example, and investigate its performance in the presence of classical channels. Denoting the average number of photons for the main signal state, in the employed efficient decoy-state protocol, by $\mu$, the secret key rate per transmitted pulse in the QKD channel, in the limit of an infinitely long key, is lower bounded by $\max [0,P({Y}_0)]$, where \cite{Lo:Decoy:2005}
\begin{equation}
P({Y}_0)=Q_{1} (1-h(e_1))-f Q_{\mu}h(E_{\mu}).
\label{Y0}
\end{equation}
Here,  $h(p)=-p{\log}_{2}p-(1-p){\log}_{2}(1-p)$ is the Shannon binary entropy function and $f$ denotes the error correction inefficiency. In (\ref{Y0}), $Q_{\mu}$, $E_{\mu}$, $Q_{1}$, and $e_1$, respectively, represent the overall gain, the quantum BER (QBER), the gain of the single photon state, and the error rate of the single photon states. 
The overall gain, $Q_{\mu}$, and the QBER, $E_{\mu}$, are, respectively, given by
\begin{equation}
Q_{\mu}=1-(1-{Y}_{0})e^{-\eta \mu}\quad \quad 
\label{Qmu}
\end{equation}
and
\begin{equation}
E_{\mu}=({Y}_{0}/2+e_{d}(1-e^{-\eta\mu})) /Q_{\mu}, 
\label{Emu}
\end{equation}
whereas the gain and the error rate of the single photon states are, respectively, as follows:
\begin{equation}
 Q_{1}={Y}_{1}\mu e^{-\mu},
\end{equation}
and 
\begin{equation}
 e_{1}=({Y}_{0}/2+e_{d}\eta)/ {Y}_{1}. 
 \label{e1}   
\end{equation}
Here, ${Y}_{0}$ represents the probability of having detector clicks at Bob's end without transmitting any photons, and ${Y}_{1}$ is the yield of a single-photon state. {Furthermore, for time-bin encoding, parameter $e_d$ models the error probability due to relative phase distortions between $r$ and $s$ pulses. The parameter $\eta$ represents the total transmissivity of the link, and is given by
\begin{equation}
\eta=\frac{1}{2}\eta_d e^{-\alpha L}.
\label{etaa}
\end{equation}	
The coefficient $1/2$ represents the loss associated with the decoder setup in Fig.~\ref{BB84}.} Denoting the repetition period of the QKD system by $T_s$, the secret key generation rate of the $m^{th}$ QKD channel is given by
\begin{equation}
\label{Eq:Rm}
R_{m}=\max [0,P({Y}_0)/T_s],
\end{equation}
where
\begin{equation}
{Y}_0=1-(1-(p_{\mathrm {dc}}+p_{m}))^2.
\label{Y0_p}
\end{equation}
In the above equation, $p_{\mathrm {dc}}=\gamma _{\mathrm {dc}} T_d$, where $\gamma _{\mathrm {dc}}$ denotes the dark count rate of a single-photon detector and $p_{m}$ denotes the total crosstalk photon count, due to Raman noise and adjacent channels, on the $m^{th}$ quantum channel, given by
\begin{equation}
p_{m}=\sum_{n=1}^{N}{(p^{\rm FR }_{nm}+p^{\rm BR}_{nm}+p^{\rm FC}_{nm}+p^{\rm BC}_{nm})}.
\label{pm}
\end{equation} 
{As explained in Sec.~II, in the DWDM structure of Fig.~\ref{single_dual}(b), $p^{\rm BR}_{nm}$ and $p^{\rm BC}_{nm}$ are both zero.} 

\section{Optimal Wavelength Assignment}
\label{Sec:Opt}
 
{Wavelength assignment in our setting can significantly affect the performance of QKD links. From Fig. \ref{raman_cross} and equations (\ref{Iforward}), (\ref{Ibackward}), (\ref{crossfor}), and (\ref{crossback}), we can infer that the crosstalk noise induced by a classical channel onto a quantum one depends on the difference between their corresponding wavelengths. Therefore, the key rate of QKD channels is dependent on the location of quantum and classical channels, with respect to each other, in the wavelength grid.} In this section, we investigate the optimal wavelength assignment that maximizes the total key rate of QKD channels, in the DWDM systems shown in Fig. \ref{single_dual}, under a minimum key rate per channel constraint. To this end, we define an optimization problem that aims to find the sets $U_{1}$, $U_{2}$, $A$, and $B$, such that the total key rate of QKD channels is maximized. This problem can be formulated as
\begin{equation}
 \max_{A,B,U_{1},U_{2} \subset G } \sum_{m=1}^{M}{R_{m}},\;\;\;\;{ \rm s.t.}\;\; R_{m}>R_{\rm th},\;\; m=1,...,M,
\label{max_rate}
\end{equation} 
where $R_{m}$ denotes the key rate of the $m^{\rm th}$ quantum channel given by (\ref{Eq:Rm}), and $R_{\rm th}$ is the minimum required value for $R_{m}$. {The parameter $R_{\rm th}$ has been defined to take into account quality-of-service considerations for QKD links. In a multi-user setup, where a minimum key rate needs to be guaranteed for each QKD user, $R_{\rm th}$ would specify this minimum rate. If we are only interested in maximizing the total key rate with no constraints on individual key rates, we can simply use a negative value for $R_{\rm th}$ in our formulation. {Given that $R_m$, by definition, is non-negative, choosing a negative value for $R_{\rm th}$ would remove any constraints on guaranteeing a minimum key rate per channel. Note that this is only for notational convenience, and otherwise a negative threshold value has no physical implications.} In the following sections, we consider both scenarios.

\begin{figure}[t]
\centering
\includegraphics[width=3 in]{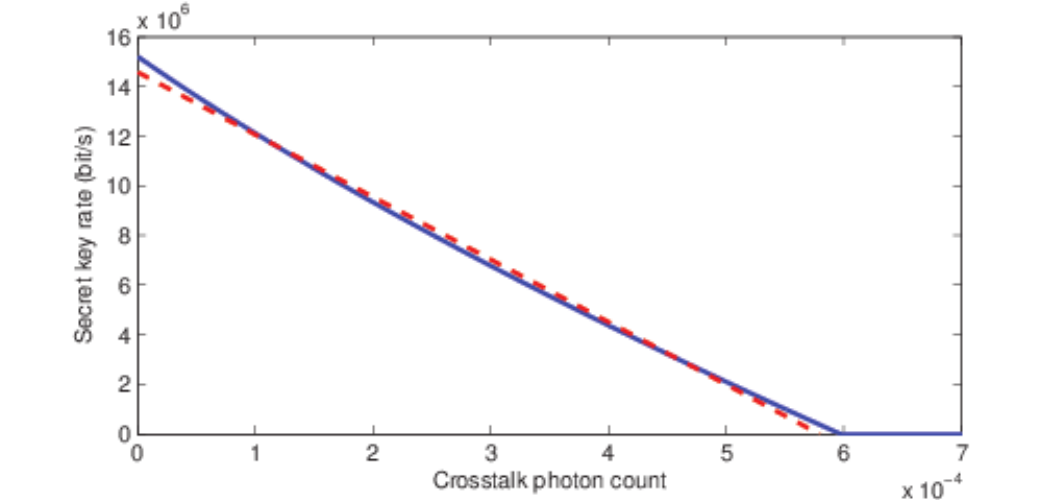}
\caption{Secret key generation rate versus {crosstalk photon count} (blue solid curve) and its linear approximation (red dashed curve).} \label{linear_rate}
\end{figure}

In order to solve the optimization problem in (\ref{max_rate}), one can simplify it by investigating the dependence of $R_m$ on $p_m$. {According to the key rate analysis presented in Sec.~\ref{Sec:Rate}, it can be concluded that $R_m$ is a descending function of $p_m$. In Appendix A, we have shown that this curve can be approximated with reasonable accuracy by a line. As an example, Fig.~\ref{linear_rate} shows the secret key rate of a single QKD channel as a function of the total crosstalk photon count, and its linear approximation, for the system parameters outlined in Table I and a fiber length of $45~{\rm km}$.} With this approximation, the optimization problem in (\ref{max_rate}) can be expressed as
\begin{equation}
\min_{A,B,U_{1},U_{2} \subset G } \sum_{m=1}^{M}{p_{m}}, \;\;\;\;{ \rm s.t.}\;\; p_{m}<p_{\rm th},\;\; m=1,...,M,
\label{min_photon}
\end{equation}
where $p_{\rm th}$ denotes the crosstalk photon count that results in $R_{m}=R_{\rm th}$ and is dependent on the fiber length and the QKD system parameters. Negative $R_{\rm th}$ values can then be modeled by infinitely large values of $p_{\rm th}$, which equivalently remove any constraints on $p_m$. In principle, the above formulation will lead to a near-optimal solution. 

In the following, we examine this optimization problem for each DWDM structure shown in Fig. \ref{single_dual}. Note that if the NBFs at the quantum receivers suppress the adjacent channel crosstalk effectively, the Raman noise will be the dominant source of crosstalk noise. Otherwise, the adjacent channel crosstalk should also be taken into account. In the following subsections, we use the criteria in (\ref{min_photon}) to find near-optimal wavelength assignments in each case. 

\begin{table}[pbt]
\caption{Nominal values used for QKD system parameters.}
\centering 
\begin{tabular}{|c |c |} 
\hline
Parameter & Value\\
\hline
Average number of photons per signal pulse, $\mu$ & 0.48\\
Quantum efficiency of single-photon detectors& 0.3\\
Receiver dark count rate, ${\gamma} _{\rm dc}$& 1E-7 ${\rm ns}^{-1}$\\
Error correction inefficiency, $f$& 1.16\\
Phase-distortion error probability, $e_d$& {0.015}\\
Laser pulse repetition interval, $T_s$ & 250 ps\\
Time gate interval, $T_d$ & 100 ps\\
Channel loss coefficient, $\alpha$ & 0.046 km${}^{-1}$\\
\hline
\end{tabular}
\end{table} 

\subsection{Full-duplex system }\label{single}
In this section, we examine the problem of wavelength assignment for the DWDM system shown in Fig.~\ref{single_dual}(a). In this case, using (\ref{pm}), the cost function in (\ref{min_photon}) can be rewritten as
\begin{eqnarray}
C=C^{b}\sum_{n=1}^{N}{\sum_{m=1}^{M}{\lambda_{q_{m}}\beta (\lambda_{b_{n}},\lambda_{q_{m}})}}+C^{f}\sum_{n=1}^{N}{\sum_{m=1}^{M}{\lambda_{q_{m}}\beta (\lambda_{f_{n}},\lambda_{q_{m}})} } \nonumber \\
+\sum_{n=1}^{N}{\sum_{m=1}^{M}{(p^{\rm FC}_{nm}+p^{\rm BC}_{nm})}}. \quad\quad\quad
\label{min_ph_ext_single}
\end{eqnarray}


\subsubsection{Raman-noise-only scenario}

In this case, we assume that the NBF used at the QKD receivers can remove the noise from adjacent channels and make them negligible. Equation (\ref{min_ph_ext_single}) {is then simplified} to
\begin{equation}
C=C^{b} Z_{1}+C^{f} Z_{2},
\label{min_ph_ext2}
\end{equation}
where
\begin{equation}
Z_{1}=\sum_{n=1}^{N}{\sum_{m=1}^{M}{\lambda_{q_{m}}\beta (\lambda_{b_{n}},\lambda_{q_{m}})}},
\end{equation}
and
\begin{equation}
Z_{2}=\sum_{n=1}^{N}{\sum_{m=1}^{M}{\lambda_{q_{m}}\beta (\lambda_{f_{n}},\lambda_{q_{m}})} }.
\end{equation}

{In the above equations, $\lambda_{b_n}\in B$ and $\lambda_{f_n}\in A$, for $n=1,...,N$. Since $C^{b}$ and $C^{f}$ are wavelength-independent, $Z_1$ and $Z_2$ should be minimized. In general, the set $A$ and $B$ can be two different sets. However, in the following, we show that in the Raman-noise-only case, the wavelengths allocated to forward and backward classical channels should be identical in the optimal scenario.
\newtheorem{first theorem}{Lemma}
\begin{first theorem}\label{noninc}
	For the optimal solutions to (\ref{min_ph_ext2}), we have $A=B$. 
\end{first theorem}
\newtheorem{nonincproof}{Proof}
\begin{nonincproof}\label{nonincproof}
	Suppose in the optimum solution $A \neq B$. Without loss of generality, assume $Z_{1}< Z_{2}$. Then, if one uses the set $A$ instead of $B$ for the backward channels, the resulting value for $Z_{2}$ would be lower. Similarly, if $Z_{1}> Z_{2}$, we can use $B$ instead of $A$ for forward channels to reduce the value of $Z_1$. This implies that $A=B$ in the optimum setting.
\end{nonincproof}
}



Using the above result, our optimization problem reduces to the case where all data channels are bidirectional. With this constraint, the optimization problem can be expressed as
{
\begin{eqnarray}
\min_{A,U_{1},U_{2} \subset G } \sum_{n=1}^{N}{\sum_{m=1}^{M}{\lambda_{q_{m}}\beta (\lambda_{f_{n}},\lambda_{q_{m}})}}, \nonumber \\
{ \rm s.t.} \;\;\sum_{n=1}^{N}{\lambda_{q_{m}}\beta (\lambda_{f_{n}},\lambda_{q_{m}})}<X_{\rm th},
\label{min_ph_opt}
\end{eqnarray}}
where $X_{\rm th}=p_{\rm th}/(C^f+C^b)$.

To solve the optimization problem in (\ref{min_ph_opt}), we propose a matrix-based algorithm. We define a $D\times D$ matrix, $\bf{P}$, with elements given by  
\begin{equation}
{\bf{P}}_{ij}=\left\{ 
\begin{array}{cc}
\lambda_{j}\beta(\lambda_{i},\lambda_{j})&\quad i\neq j\\
\infty&\quad i=j
\end{array}
\right.
.
\label{linear_opt_prob}
\end{equation}

\begin{algorithm}[!t]
		\caption{Near-Optimal Wavelength Assignment Algorithm}\label{opt_algorithm}
		\begin{algorithmic}[4]
			\Statex {{\textbf{Input:}} $\bf{P}$, $M$, $N$, $D$, $X_{th}$}
			\Statex {{\textbf{Output:}}}
			\Statex {{Vector ${\bf q}$} containing the indices of the elements of G assigned to quantum channels } 
			\Statex {{Vector ${\bf c}$} containing the indices of the elements of G assigned to classical channels } 
			\If {$\left({D \above 0pt N}\right)\leq \left({D \above 0pt M}\right)$}
			\State{$\bf{Z}=\bf{P}$}
			\State{${\bf{A}}=$ matrix of all size-$N$ subsets of $\{1\ldots D\}$; each row represents one valid subset	}
			\Else
			\State{$\bf{Z}=\bf{P^{T}}$}
			\State{${\bf{A}}=$ matrix of all size-$M$ subsets of $\{1\ldots D\}$; each row represents one valid subset	}
			\EndIf		
			\State{$t=1000$}
			\For{$i=1: \min\left(\left({D \above 0pt N}\right), \left({D \above 0pt M}\right)\right)$}
			\State{${\bf{b}}={\bf{A}}(i,:)$}
			\State{${\bf{y}}=\sum_{j \in {\bf{b}} }{{\bf{Z}}(j,1:D)}$}
			\State{$[{\bf{d}}, {\bf{index}}]={\rm sort}({\bf{y}})$}
			\If {$\left({D \above 0pt N}\right)\leq \left({D \above 0pt M}\right)$} 
			\State{$s=\sum_{j=1}^{M}{d_{j}}$}
			\If {$s<t \;$ and ${\bf{d}}(M)<X_{th}$}
			\State{$t=s$}
			\State{${\bf q}={\bf{index}}(1:M)$}
			\State{${\bf c}=\bf{b}$}
			\EndIf
			\Else
			\State{$s=\sum_{j=1}^{N}{d_{j}}$}
			\State{${\bf{w}}=\sum_{j \in {\bf{index}}(1:N)}{{\bf{Z}}({\bf{b}},j)}$}
			\If {{$s<t \;$} and $\max({\bf{w}})<X_{th}$}
			\State{t=s}
			\State{${\bf q}=\bf{b}$}
			\State{${\bf c}={\bf{index}}(1:N)$}
			\EndIf
			\EndIf
			\EndFor
		\end{algorithmic}
\end{algorithm}
{The elements of $\bf{P}$ are defined based on the summands in (\ref{min_ph_opt}). $\bf{P}_{ij}$ represents the Raman cross section corresponding to $\lambda_{i}$, as the data channel wavelength, and $\lambda_{j}$, as the quantum channel wavelength, multiplied by $\lambda_{j}$. Since classical and quantum channels have different wavelengths, we have chosen $\bf{P}_{ij}=\infty$ for $i=j$. The optimization problem in (\ref{min_ph_opt}) can be interpreted as finding $N$ rows and $M$ columns of matrix $\bf{P}$ such that the summation of elements at the intersection of these rows and columns is minimum (the diagonal elements of $\bf{P}$ will automatically be excluded because of their infinitely high value), and the constraint in (\ref{min_ph_opt}) is satisfied. The proposed optimization algorithm is presented in Algorithm 1, which is self explanatory. It should only be mentioned that in line 12, $\bf{d}$ is the sorted version of $\bf{c}$ in the ascending order, and $\bf{index}$ is the vector of corresponding indices. This matrix-based algorithm is also applicable to other scenarios we consider in our work. In each case, we just need to find the relevant matrix $\bf{P}$ and apply Algorithm 1 to it.


\subsubsection{Raman + Adjacent channel crosstalk scenario}

In this case, we consider all the terms in (\ref{min_ph_ext_single}) to determine the optimum wavelength pattern. Note that $p^{\rm FC}_{nm}$ and $p^{\rm BC}_{nm}$ are only present for $|\lambda_{f_n}-\lambda_{q_m}|=\Delta$ and $|\lambda_{b_n}-\lambda_{q_m}|=\Delta$, respectively. Hence, we use the results of the previous case and propose a suboptimal wavelength assignment method that assumes bidirectional data channels. With this constraint, (\ref{min_ph_ext_single}) reduces to
\begin{equation}
 C=\sum_{n=1}^{N}{\sum_{m=1}^{M}{\left\{(C^{b}+C^{f})\lambda_{q_{m}}\beta (\lambda_{f_{n}},\lambda_{q_{m}})+p^{\rm BC}_{nm}+p^{\rm FC}_{nm}\right\}}}.
 \label{ramanadj_single}
\end{equation}
This problem can also be solved by Algorithm 1. In this case, the elements of matrix $\bf{P}$ are given by
 \begin{equation}
{\bf{P}}_{ij}=\left\{ 
\begin{array}{cc}
(C^{f}+C^{b})\lambda_{j}\beta(\lambda_{i},\lambda_{j})+p^{\rm BC}_{ij}+p^{\rm FC}_{ij}&\quad i\neq j\\
\infty&\quad i=j
\end{array}
\right. ,
\label{linear_opt_prob_2}
\end{equation}
and $X_{\rm th} = p_{\rm th}$.

\subsection{Dual-fiber system}

Now let us consider the dual-fiber system in Fig.~\ref{single_dual}(b). In this scenario, the optimization problem in (\ref{min_photon}) can be split into two problems, one for each employed fiber. In this case, the number of classical channels per fiber is fixed to $N$ but we have to decide how many, out of $M$, quantum channels need to be allocated to each fiber. Suppose $M=1$. Then, it does not matter which fiber we use for the QKD channel, but we need to find the optimum wavelength assignment that minimizes the crosstalk noise. This way we find $\lambda_{q_1}$. Now, if $M=2$, we can use the same wavelength assignment but on the other fiber and the total key rate is expected to be higher than the case where both QKD channels are on the same fiber. We can keep adding QKD channels to the game, but it can be seen that the optimum assignment should have $k=\left\lfloor M/2\right\rfloor$ QKD channels on one fiber and $M-k$ channels on the other. Now that we have a fixed number of QKD channels on each link, we can solve the two optimization problem, corresponding to forward and backward links, separately. Using (\ref{pm}), the cost function in (\ref{min_photon}) for each optimization problem is given by
\begin{equation}
C=C^{f}\sum_{n=1}^{N}{\sum_{m\in S}{\lambda_{q_{m}}\beta (\lambda_{f_{n}},\lambda_{q_{m}})} }
+\sum_{n=1}^{N}{\sum_{m\in S}{p^{\rm FC}_{nm}}},
\label{min_ph_ext_dual}
\end{equation}  
where $S=\{1,...,k\}$ for the forward fiber link, and $S=\{k+1,...,M\}$ for the backward one. Note that if $M$ is an even number, $k=M/{2}$. In this case, the two optimization problems are identical and achieve similar wavelength assignment patterns.

\subsubsection{Raman-noise-only scenario}

 In this case, the cost function in (\ref{min_ph_ext_dual}) reduces to
 \begin{equation}
 C=\sum_{n=1}^{N}{\sum_{m\in S}{\lambda_{q_{m}}\beta (\lambda_{f_{n}},\lambda_{q_{m}})} }.
 \label{min_ph_opt_dual_m}
 \end{equation}
 Comparing (\ref{min_ph_opt_dual_m}) and (\ref{min_ph_opt}), it is concluded that this optimization problem can be solved by Algorithm 1 for $M=k$ (and $k+1$ for odd values of $M$), with the matrix $\bf{P}$ described in (\ref{linear_opt_prob}), and the threshold $X_{\rm th}=p_{\rm th}/C^{f}$.
 

\subsubsection{ Raman + Adjacent channel crosstalk scenario}
 
In this case, the optimization problem in (\ref{min_ph_ext_dual}) should be solved. Here again, Algorithm 1, for $M=k$ (and $k+1$ for odd values of $M$), at $X_{\rm th} = p_{\rm th}$, can be used with matrix $\bf P$ defined as
\begin{equation}
{\bf{P}}_{ij}=\left\{ 
\begin{array}{cc}
C^{f}\lambda_{j}\beta(\lambda_{i},\lambda_{j})+p^{\rm FC}_{ij}&\quad i\neq j\\
\infty&\quad i=j
\end{array}
\right.
.
\label{linear_opt_prob_3}
\end{equation}
  
\section{Numerical Results} \label{Sec:Num}

\begin{table}[pbt]
\caption{Nominal values used for the DWDM system parameters.}
\centering 
\begin{tabular}{|c |c |} 
\hline
Parameter & Value\\
\hline
Channel spacing, $\Delta$&  200 GHz\\
{Adjacent channel isolation of DWDM module, $\gamma_a$} & 30 dB\\
{Directivity of DWDM module, $\chi_a$}& 50 dB\\
Bandwidth of NBF, $\Delta \lambda$& 15, 125 GHz\\
\hline
\end{tabular}
\end{table} 

In this section, the proposed wavelength assignment methods are investigated in more detail. Our example DWDM system uses the wavelength grid ranging from $1530~{\rm nm}$ to $1565~{\rm nm}$ in the C-band with a nominal 0.2~dB/km loss across the grid (corresponding to $\alpha = 0.046$/km). The nominal values for QKD systems are listed in Table I, and other system parameters are summerized in Table II. These parameters are chosen based on certain practical considerations. {We assume that, in the full-duplex DWDM system, the classical channels use on-off keying with the data rate of $10~{\rm GHz}$. The launch power of the data laser is controlled by the receiver sensitivity, which is assumed to be $-28~{\rm dBm}$, corresponding to a bit error rate of $10^{-12}$. As for the dual-fiber structure, we assume that $100{\rm G}$ coherent systems are used in the data links. The power of the received classical signal in both structures is chosen to be $-25~{\rm dBm}$.} We consider different cases of ``Raman noise only" and ``Raman + Adjacent channel crosstalk", based on the bandwidth of the NBF used, for full-duplex and dual-fiber DWDM systems. We consider two cases for the bandwidth of the NBF at the quantum receivers: $15~{\rm GHz}$, and $125~{\rm GHz}$. We assume that in the first case, the adjacent channel crosstalk is suppressed effectively so that it can be neglected. As for the $125~{\rm GHz}$ NBF, we assume that a Gaussian shaped filter is used, which causes an attenuation of about $16~{\rm dB}$ at the passband of adjacent channels. 

In order to obtain the Raman cross section $\beta (\lambda_{d},\lambda_{q})$ in (\ref{linear_opt_prob}), (\ref{linear_opt_prob_2}) and (\ref{linear_opt_prob_3}) for different values of $\lambda_{d}$ and $\lambda_{q}$, we use the measurement results shown in Fig.~\ref{raman_cross}. The results in Fig.~\ref{raman_cross} are, however, for the specific case of $\lambda_{d} = 1550$~nm. In order to use the same measurement results for an arbitrary $\lambda_{d}$, we use two tricks. First, we find wavelength $\lambda_{\delta}$ such that  
\begin{equation}
\frac{1}{1550{\rm \ nm}}-\frac{1}{\lambda_\delta}=\frac{1}{\lambda_d}-\frac{1}{\lambda_q},
\end{equation}	
that is, the frequency difference between $\lambda_{\delta}$ and 1550~nm is the same as that of $\lambda_{q}$ and $\lambda_d$. $\beta (\lambda_{d},\lambda_{q})$ is then expected to be proportional to $\beta(1550 {\rm \ nm},\lambda_\delta)$, which can be obtained from Fig.~\ref{raman_cross}. Given that the Raman cross section is known to be proportional to $(1/\lambda_{q})^4$ as well \cite{islam2007raman}, we assume that 
\begin{equation}
\beta (\lambda_{d},\lambda_{q})=(\frac{\lambda_\delta}{\lambda_q})^4 \beta(1550 {\rm \ nm},\lambda_\delta).
\end{equation}
Our numerical results indicate that the term in power 4 has little effect on our final results.

In the following, our numerical results are presented. We investigate the performance of our proposed wavelength assignment methods in terms of their key rate enhancement and optimality. Furthermore, the wavelength patterns obtained by these schemes are examined.

\subsection{Rate Enhancement}
 \begin{figure}[t]
 	\centering
 	\includegraphics[width=14 cm]{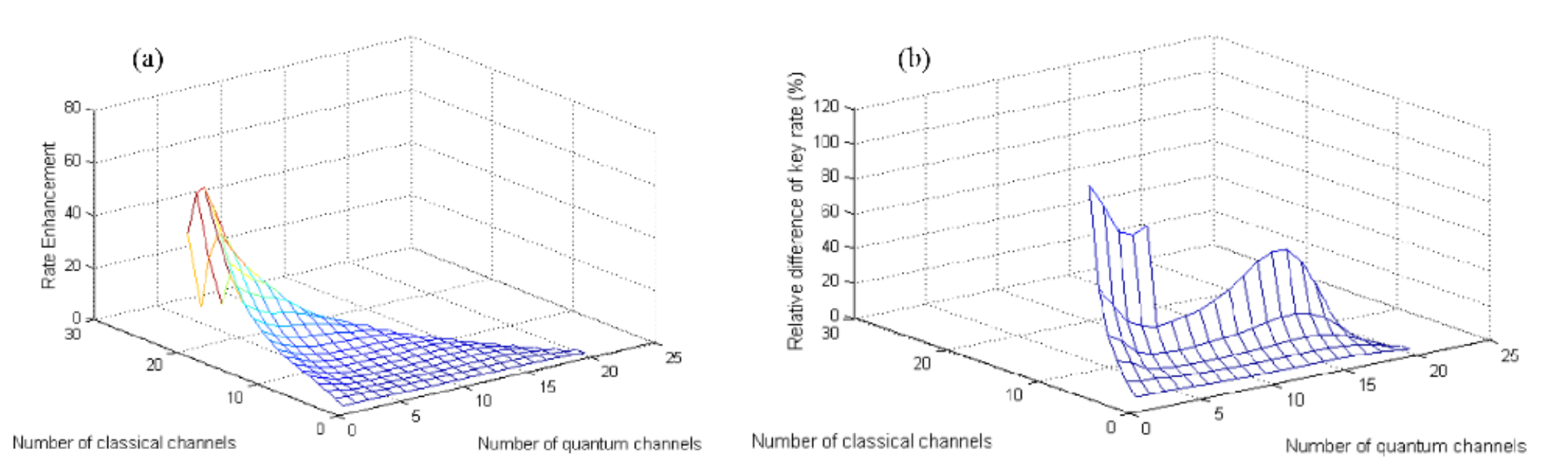}
 	\caption{Rate enhancement at (a) $L=50~{\rm km}$ and (b) $L=65~{\rm km}$ for the full-duplex system at $R_{\rm th}=0$.}  \label{conv} 
 \end{figure}

In this section, we compare our proposed methods with the conventional approach of assigning the lower part of the wavelength grid to the quantum and the longer wavelengths to the classical channels. We define a rate enhancement measure, denoted by $RE$, as follows:
\begin{equation}
RE=\frac{R_{\rm pr}-R_{\rm co}}{R_{\rm co}}\times 100,
\end{equation}
where $R_{\rm pr}$ and $R_{\rm co}$ are the total key rate obtained by the proposed and the conventional methods, respectively. 

First, we consider the case of ``Raman noise only" for the full-duplex system. We choose $R_{\rm th}=0$, i.e, all quantum channels are required to have a positive key rate. The rate enhancement parameter, $RE$, for different values of $M$ and $N$, is shown in Figs.~\ref{conv}(a) and (b), for $L=50~{\rm km}$ and $L=65~{\rm km}$, respectively. It can be seen that, our proposed method can improve the total key rate significantly, especially for large $N$ and small $M$. The rate can be improved by over 100\%, as shown in Fig.~\ref{conv}(b), if we are in a region that the system is sensitive to the amount of the background noise. At $L=65$~km, the channel loss is higher, hence the resilience of the QKD system to the background noise would be lower than that of $L = 50$ km. {To further investigate the rate enhancement at different fiber lengths, Table~\ref{T3} summarizes the secret key rate of the proposed and conventional methods for $N=12$ and $M=1$. It can be seen that, as fiber length increases, the rate enhancement increases as well. In particular, at $L=60~{\rm km}$, while the key rate of the conventional approach is zero, we can still obtain positive secret key rates by using our proposed method. This implies that our near-optimal wavelength assignment technique could increase the maximal security distance of QKD systems.

\begin{table}[pbt]
	\caption{Secret key rate of the proposed and conventional methods for $N=12$ and $M=1$ at different fiber lengths. \label{T3}
}
	\centering 
	\begin{tabular}{|c |c | c | c |c| c| c| } 
		\hline
		Fiber length (km) & 40 & 45 & 50 & 55 & 60 & 65 \\
		\hline
		$R_{\rm pr}$ (bit/s)&  1.49E7 & 1.02E7 & 6.29E6 & 2.93E6 & 4.5E4 & 0\\
		\hline
		$R_{\rm co}$ (bit/s) & 1.41E7 & 9.33E6 & 5.27E6 & 1.79E6 & 0 & 0 \\
		\hline
		RE ($\%$)&  5.5 & 9.69 & 19.35 & 63 & $\infty$ & 0\\
		\hline
	\end{tabular}
\end{table} 	
} 

In order to further investigate the performance of our proposed methods, we define another measure, denoted by $N_{\rm max}$, as the maximum possible number of classical channels that can be integrated with $M$ quantum channels such that all of them have a positive key rate. We compare this parameter for the proposed and the conventional methods. Our numerical results show that, depending on the fiber length, $N_{\rm max}$ can often be improved by one or two channels. This means that by the use of optimal wavelength assignment higher data traffic can be supported.

We have also considered other cases of ``Raman noise only" for dual-fiber systems and ``Raman + Adjacent channel crosstalk" for full-duplex and dual-fiber DWDM systems. Our numerical results show that the conclusions drawn in this section can be extended to these cases as well.    
  
\subsection{Near-optimal Wavelength Patterns}
 \begin{figure}[t]
 	\centering
 	\includegraphics[width=17 cm]{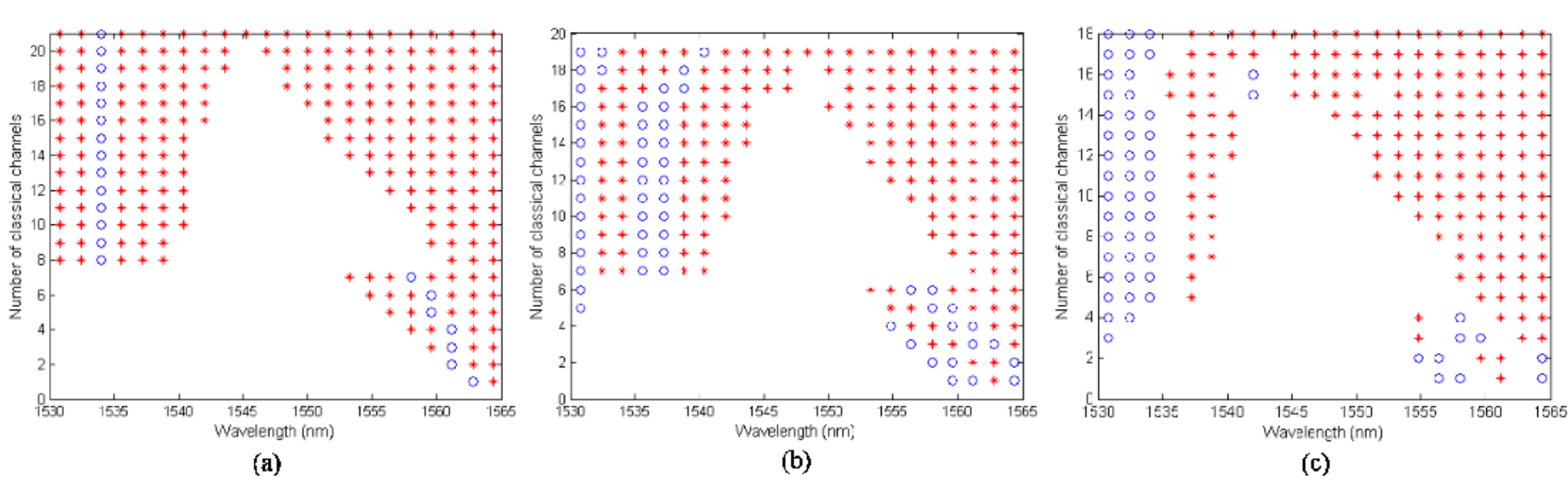}
 	\caption{Proposed wavelength assignment patterns in a 200 GHz full-duplex system in (a) ``Raman noise only" case for $M=1$, (b) ``Raman noise only" case for $M=3$, and (c) ``Raman + Adjacent channel crosstalk" case for $M=3$ at $L=50~\rm{km}$. Each row depicts the optimum location of quantum and classical channels in the wavelength grid, where {\color{red} ${\ast}$} represents a classical channel and ${\color{blue}\circ}$ represents a quantum one.} \label{waveassign}
 \end{figure}

In this section, the wavelength assignment patterns obtained by the proposed near-optimal methods are investigated. We assume that $R_{\rm th}<0$, which corresponds to the case that the total key rate is maximized with no constraint on the individual key rates. We particularly look at the cases where substantial gain can be achieved by optimizing the wavelength assignment, i.e., when a small number of quantum channels are present. Here, we choose $M=1$ and $M=3$ and examine the wavelength patterns, first, in the Raman-noise only case. Figures \ref{waveassign}(a) and (b) depict the proposed wavelength assignment for, respectively, $M=1$ and $M=3$ quantum channels and different values of classical channels, $N$, for a 200 GHz full-duplex DWDM system. Note that these figures also provide the proposed wavelength assignment for each individual fiber in a dual-fiber system for, respectively, $M=2$ and $M=6$ quantum channels. In each figure, each row shows the proposed locations for the quantum, represented by ``$\circ$'', and classical, represented by ``$\ast$'', channels for each given number of classical channels. As can be seen, the proposed pattern for each $N$ is not necessarily compatible with the conventional method of having two separate quantum and classical bands at the two ends of the wavelength grid even if we only have one quantum channel. This pattern is, in general, consisted of multiple interspersed quantum and classical bands. For example, in Fig.~\ref{waveassign}(a), we can see that the QKD channel is between two classical bands for $N\geq 3$. This result can be explained by referring to Fig.~\ref{raman_cross}. According to this figure, the Raman noise takes its smallest values in regions $S_1$ and $S_2$, in Fig.~\ref{raman_cross}, on the two sides of the pump laser wavelength. In the Raman-noise only scenario, our optimum wavelength assignment, then benefits from this low-noise regions to improve performance. {For instance, in the case of $N=1$, where the classical channel is assigned to 1564.4~nm, the Raman noise at 1562.8~nm is 2.2E-5, while it is 3.2E-5 at 1530.8~nm. In the case of $N=2$, where the two classical channels are assigned to 1564.4~nm and 1562.8~nm, the Raman noise at 1561.2 nm is 4.46E-5, while it is 6.48E-5 at 1530.8 nm.}

Next, we investigate the wavelength assignment patterns in the ``Raman + Adjacent channel crosstalk" case for the full-duplex system. The proposed wavelength assignment for $M=3$ is shown in Fig.~\ref{waveassign}(c). As can be seen, the wavelength assignment patterns are different from the "Raman noise only" case. With the chosen system parameters, the adjacent channel crosstalk can be more than the Raman noise. Hence, the wavelength assignment method avoids the allocation of a quantum and a classical channel to adjacent wavelengths. This would result, especially when $N$ is large, in the optimum solution converging to the conventional one as can be seen for $N>16$ in Fig.~\ref{waveassign}(c). More generally, our results imply that if the adjacent channel crosstalk is the dominant source of noise and $R_{\rm th}<0$, the wavelength assignment pattern converges to the conventional method solution, when the capacity of the system is almost fully used. 

It is interesting to study the dependence of the optimal wavelength pattern on the transmission distance, or, effectively, the channel loss. In our formulation, the key parameters that are distance dependent are $C^{f}$ and $C^{b}$, which affect the cost function, as well as $R_m$, whose value must satisfy our optimisation constraint $R_m > R_{\rm th}$. In all cases that there are no constraints on the key rate, i.e., when $R_{\rm th} < 0$, the latter dependence on the distance does not matter. As for the former, it turns out that in the “Raman-noise-only” scenarios, the cost function can be made independent of the fiber length by eliminating $C^{f}$ and $C^{b}$ as in (\ref{min_ph_opt}) and (\ref{min_ph_opt_dual_m}). Hence, for $R_{\rm th} < 0$, the achieved wavelength assignment patterns in the Raman-noise-only scenarios, e.g. the result in Fig.~\ref{waveassign}(a) and (b), would be independent of transmission distance. In all other cases, the optimal wavelength pattern can, in principle, depend on the channel loss. However, in the particular cases we have considered for our numerical results, e.g. in Fig.~\ref{waveassign}(c), we have verified that up to $L=120~\rm{km}$ the obtained wavelength patterns remain the same.

\subsection{Optimality}

In this section, we compare the proposed near-optimal method, based on (\ref{min_photon}), with the optimum approach that maximizes the total key rate in (\ref{max_rate}). { We have found the solution to the latter by an exhaustive search.} First, we choose $R_{\rm th}<0$. In this case, our numerical results show that, for low fiber lengths, e.g., $L=45~\rm{km}$, the proposed methods generally lead to the same wavelength assignment patterns, hence, the same total key rates, that the optimum solution offers. There are a few exceptions. However, even in those few cases, the percentage of the relative difference of the total key rate is below $0.001\%$. 

As the fiber length increases, and for sufficiently large values of $M$ and $N$, the proposed and the optimum solutions may lead to different total key rates. The reason is mainly because of the linear approximation we use to convert maximizing the key rate criterion into minimizing the background noise. At large distances, QBER for some channels may be very large resulting in zero key rates for them. In fact, when $M$ is sufficiently large, the optimum solution may include some quantum channels with zero key rates. Our noise-based solution will instead try to distribute the noise almost equally among all channels, which, in certain cases, would result in lower total key rates than that of the optimum solution. As an example, Fig.~\ref{rate_opt} shows the total key rate for $N=8$ classical channels at $L=62~\rm{km}$ for different values of $M$. As can be seen, when $R_{\rm th}<0$, for $M>5$ the optimum solution remains the same because new QKD channels will have zero key rates. Our proposed method, however, achieves a lower total key rate by supporting a larger number of users with positive key rates. It can be concluded that, for each $N$ and $L$, there is a specific value of $M$ for which the maximum total key rate is achieved. Increasing $M$ beyond that value does not increase the total key rate and only quantum channels with zero key rates are added. This is an important observation if maximizing the total key rate is the key objective of the operator.

Next, we consider the case of $R_{\rm th}=0$, i.e., when we need to guarantee a positive key rate for each quantum channel. From Fig.~\ref{rate_opt}, it can be seen that, the proposed method has a reasonable accuracy in this case, since the linear approximation method used is more accurate in this positive-key region. Here, we again see that the price of supporting a larger number of quantum users could be a lower total key rate. It depends on which criterion, number of users versus total key rate, is more important for the operator in order to decide on the right working point. Our analysis, in any case, enables the operators to plan wisely for their resource allocation.

    \begin{figure}[t]
    	\centering
    	\includegraphics[width=3 in]{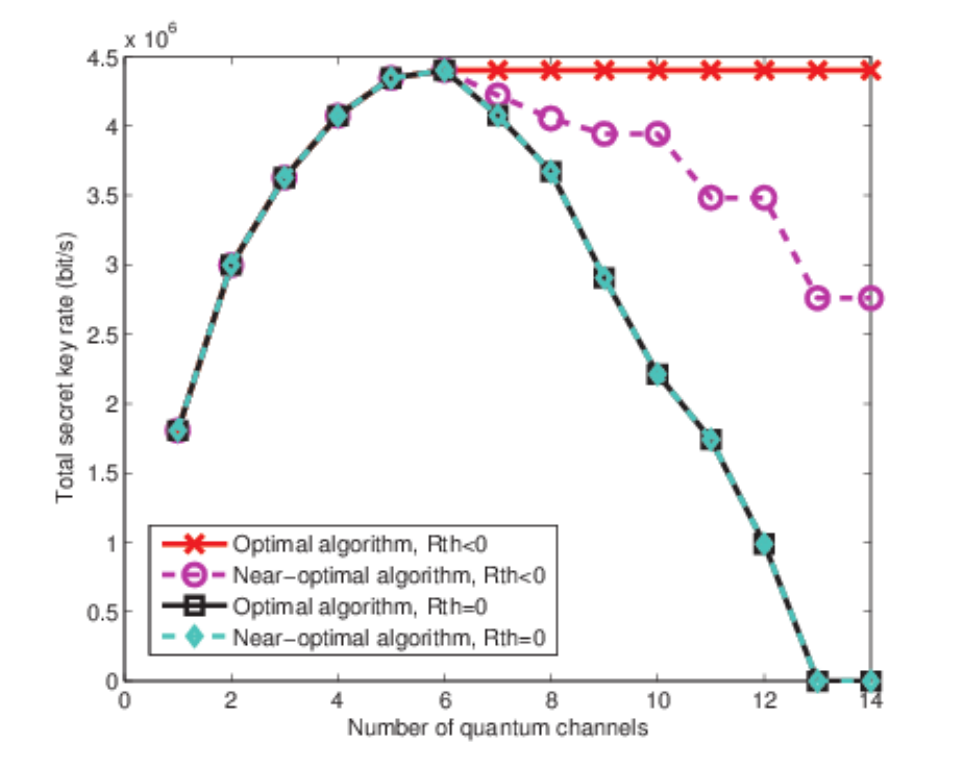}
    	\caption{Comparison between the total key rate of the optimal and near-optimal methods for $N=8$ and $L=62~{\rm km}$.} \label{rate_opt}
    \end{figure}

\section{Conclusions}
\label{Sec:Conc}
In this paper, we considered the problem of wavelength assignment in a hybrid quantum-classical network. We exploited the reconfigurability of optical networks to improve the performance of QKD links by appropriate wavelength allocation. We considered two system setups, namely, full-duplex and dual-fiber, and investigated the optimum wavelength assignment method that maximizes the total key rate of QKD channels in each case. In our analysis, two main sources of crosstalk, Raman noise and adjacent channel crosstalk were considered. We used linear approximations to propose efficient near-optimal wavelength assignment methods for these cases. Furthermore, various simulations and numerical investigations were carried out to examine the proposed methods. Our numerical result showed that the conventional wavelength assignment method of two separate quantum and classical bands would not necessarily be the optimum solution. Instead, the optimal wavelength allocation pattern could include several quantum and classical bands interspersed among each other. We showed that in most cases our proposed wavelength assignment methods were nearly identical to the optimum method. Furthermore, we showed that our proposed method could significantly improve the total key rate of the DWDM system especially in the noise dominated regimes. We found that for any given number of classical channels, there would exist an optimum number of quantum channels for which the total key rate would be maximum. It is worth mentioning that since our proposed methods minimize the total crosstalk noise, they can be used for different QKD protocols.

\begin{appendices}
	\section{Linear Approximation of Secret Key Rate}
	In this appendix, we analyze the secret key rate of a quantum channel and derive a linear approximation to it in certain regions of interest. From equation (11) in the main text, we can write
	\begin{equation}
	P({Y}_0)=Q_{1} x_{1}-f Q_{\mu}x_{2},
	\label{Y0_app}
	\end{equation}
	where
	\begin{eqnarray}
	\label{eq:x}
	x_{1}(e_1)=1-h(e_1),\nonumber \\
	x_{2}(E_{\mu})=h(E_{\mu}).
	\end{eqnarray}
	{To the first-order approximation, and for sufficiently small values of $p$, the entropy function $h(p)$ can be approximated as a linear function of $p$. In our case, the relevant values for $e_1$ and $E_\mu$ in (\ref{eq:x}) are expected to be small if we want to have positive key rates. For instance, if we assume that $\eta_d=1$, $L=0$, $e_d=0$, and $f=1$, and just increase $Y_0$ until $P(Y_0)$ becomes zero, we find that $e_{1}<E_{\mu}<0.0953$. That is, if we restrict ourselves to the regime in which key rates are positive, we can approximate $x_1$ by $a e_1 + b$ and $x_2$ by $k E_\mu + j$, for some constant parameters $a$, $b$, $k$, and $j$. We have verified that under the assumption of $e_{1}<E_{\mu}<0.0953$, the mean square error for these approximations is less than $1.89E-4$. Now, if we substitute these linear approximations into (\ref{Y0_app}) and use equations (12)-(15) in the main text, we obtain 
		\begin{equation}
		P({Y}_0)=UY_{0}+V,
		\end{equation}
		where}
	\begin{equation}
	U=\frac{a}{2} \mu e^{-\mu}+b(1-\eta)\mu e^{-\mu}-\frac{k}{2} f-f j e^{-\eta \mu},
	\end{equation}  
	and 
	\begin{equation}
	V=a \eta e_{d} \mu e^{-\mu}+b\eta\mu e^{-\mu}-k f e_{d}(1-e^{-\eta \mu})-f j (1-e^{-\eta \mu}). 
	\end{equation} 
	Finally, noting that, for $(p_{\rm dc}+p_{m})\ll 1$, equation (18) in the main text can be approximated by ${Y_0}\simeq 2p_{\rm dc}+2p_{m}$, we obtain
	\begin{equation}
	P(Y_{0}) \approx 2U p_{m}+ 2U p_{\rm dc}+V,
	\end{equation}
	which means that the key rate can be written in a linear form versus $p_m$. Obviously, the above approximation holds when $p_m$ is small enough that we still get positive key rates. Once this condition breaks down, as we saw in Fig.~7 in the main text, the optimal solution that maximizes the total key rate can differ from the one that minimizes the background noise.
\end{appendices}

\bibliographystyle{IEEEtran}

\end{document}